\documentclass[12pt]{article} 
\usepackage{amsmath}
\usepackage{amssymb} 
\usepackage{amscd}
\usepackage{amsthm}
\usepackage[utf8]{inputenc}
\usepackage{stmaryrd}
\usepackage{hyperref}
\def\dim{{\mbox{dim}}}

\def\calm{{\cal M}}

\def\calh{{\cal H}}

\def\calc{{\cal C}}

\def\bbbone{\mbox{\rm 1\hspace {-.6em} l}}

\def\Diff{{\mbox{Diff}}}
\def\Aut{{\mbox{Aut}}}

\numberwithin{equation}{section}

\begin{document}

\enlargethispage{3cm}

\thispagestyle{empty}
\begin{center}
{\bf EXCEPTIONAL QUANTUM GEOMETRY }
\end{center} 
\begin{center}
{\bf AND PARTICLE PHYSICS II}
\end{center}
\vspace{1cm}

\begin{center}
 Michel DUBOIS-VIOLETTE
\footnote{Laboratoire de Physique Th\'eorique,
 CNRS, 
Universit\'e Paris-Sud,
Universit\'e Paris-Saclay,
B\^atiment 210,
F-91 405 Orsay.\\
Michel.Dubois-Violette$@$u-psud.fr}
and
Ivan TODOROV
\footnote{INRNE,
Bulgarian Academy of Sciences,
Tsarigradsko Chaussee 72, BG-1784
 Sofia.\\
 ivbortodorov@gmail.com}
\end{center}

\vspace{2cm}

\begin{abstract}

We continue the study undertaken in \cite{mdv:2016b} of the relevance of the exceptional Jordan algebra $J^8_3$ of hermitian $3\times 3$ octonionic matrices for the description of the internal space of the fundamental fermions of the Standard Model with 3 generations. By using the suggestion of  \cite{tod-dre:2018} (properly justified here) that the Jordan algebra $J^8_2$ of hermitian $2\times 2$ octonionic matrices is relevant for the description of the internal space of the fundamental fermions of one generation, we show that, based on the same principles and the same framework as in \cite{mdv:2016b}, there is a way to describe the internal space of the 3 generations which avoids the introduction of new fundamental fermions and where there is no problem with respect to the electroweak symmetry.

\end{abstract}
\vfill

\noindent LPT-ORSAY 18-78 
\newpage

\tableofcontents

\newpage

\section{Introduction}
It is natural to associate the finite spectrum of fundamental particles to a finite quantum space endowed possibly with some additional structure. This point of view underlies, with some variations, most noncommutative geometric approaches  to particle physics (see e.g. \cite{mdv-ker-mad:1989a}, \cite{mdv-ker-mad:1988b}, \cite{mdv:1991}, \cite{ac-lot:1990}, \cite{cha-ac:1997}, 
\cite{cha-ac-mar:2007}). Such a finite quantum space is a virtual space associated to its algebra of observables that is a finite-dimensional real algebra which is a quantum analog of a finite-dimensional algebra of real functions.\\

In \cite{mdv:2016b} an argument based on an interpretation of the quark-lepton symmetry in terms of the unimodularity of the color group and the existence of 3 generations was presented suggesting that the finite quantum space corresponding to the exceptional Jordan algebra $J^8_3=\calh_3(\mathbb O)$ of $3\times 3$ hermitian octonionic matrices is relevant for the description of the internal space of the 3 generations of fundamental fermions. The triality underlying the structure of $J^8_3$ corresponds to the 3 generations while the representation of the octonion algebra $\mathbb O$ as a complex 4-dimensional space $\mathbb C\oplus \mathbb C^3$ is associated with the quark-lepton symmetry ($\mathbb C^3$ for the quark and $\mathbb C$ for the corresponding lepton). This part of 
\cite{mdv:2016b} will be briefly reviewed in Sections 2 and 3.\\

The quark-lepton symmetry means here (as in \cite{mdv:2016b}) the fact that to each quark corresponds one and only one lepton.\\

The automorphism group of the octonion algebra $\mathbb O$ is the exceptional group $G_2$ while the subgroup which preserves the splitting $\mathbb C\oplus \mathbb C^3$ is $SU(3)$ which in our interpretation should be identified to the color group $SU(3)_c$ (compare to the approach of \cite{gun-gur:1973}). The automorphism group of the exceptional algebra $J^8_3$ is the exceptional group $F_4$ while the subgroup which preserves the representation of the octonions occurring in the elements of $J^8_3$ as elements of $\mathbb C\oplus \mathbb C^3$ is isomorphic to $SU(3)\times SU(3)/\mathbb Z_3$ where the first factor $SU(3)$ should be again identified to the color group $SU(3)_c$  (see  Section 3 and  \cite{mdv:2016b} and Section 2.12 of \cite{yok:2009}).\\

In \cite{tod-dre:2018} it has been suggested that the quantum space corresponding to the Jordan subalgebra $J^8_2=JSpin_9$ of $J^8_3$ is relevant for the description of the fundamental fermions of the Standard Model (i.e. of one generation). Here $J^8_2$ is the Euclidean Jordan algebra $\calh_2(\mathbb O)$ of hermitian $2\times 2$-matrices with octonionic entries. The automorphism group of $J^8_2$ is $Spin_9$ and the intersection in $F_4$ of $Spin_9$ with $SU(3)\times SU(3)/\mathbb Z_3$  is precisely the group of symmetry $G_{SM}=SU(3)\times SU(2)\times U(1)/\mathbb Z_6$ of the Standard Model, where $SU(3)$ above is again the color group $SU(3)_c$ (see \cite{tod-mdv:2017}). This latter point has an intrinsic meaning in terms of $J^8_2$, namely $G_{SM}=SU(3)\times SU(2)\times U(1)/\mathbb Z_6$ is the subgroup of the automorphism group $Spin_9$ of $J^8_2$ which preserves the representation of the octonions occurring in the elements of $J^8_2$ as elements of $\mathbb C\oplus \mathbb C^3$, (i.e. which preserves the imaginary unit of $\mathbb O$ playing the role of $i\in \mathbb C$). This part is the content of Section 4.\\

In Section 5 it is noticed that corresponding to the triality there are 3 canonical Jordan subalgebras  $J_{(1)}, J_{(2)}, J_{(3)}$ of $J^8_3$ isomorphic to $J^8_2=JSpin_9$ associated to the 3 diagonal primitive idempotents
\[
E_{(1)}=\left(
\begin{array}{ccc}
1 & 0 & 0\\
0 & 0 & 0\\
0 & 0 & 0
\end{array}
\right),\>\>
E_{(2)}=\left(
\begin{array}{ccc}
0 & 0 & 0\\
0 & 1 & 0\\
0 & 0 & 0
\end{array}
\right),\>\>
E_{(3)}=\left(
\begin{array}{ccc}
0 & 0 & 0\\
0 & 0 & 0\\
0 & 0 & 1
\end{array}
\right)
\]
defined by $J_{(i)}=(\bbbone - E_{(i)})\> J^8_3(\bbbone-E_{(i)})$. This leads us to a correspondence triality-generations slightly different conceptually from the one suggested in Section 4.3 of \cite{mdv:2016b} ;  here there is no potential problem with respect to the electroweak symmetry and no additional new Majorana fermions.\\

Indeed the automorphism groups $Aut(J_{(i)})$ of the $J_{(i)}$ are 3 subgroups of $F_4$ which are isomorphic to $Spin_9$, they are the subgroups of $F_4$ leaving respectively invariant the idempotents $E_{(i)}$ $(i\in \{1,2,3\})$. The intersections $G^{(i)}_{SM}$ of $Aut(J_{(i)})$ with the subgroup $SU(3)\times SU(3)/\mathbb Z_3$ of $F_4$ are 3 subgroups $G^{(i)}_{SM}$ isomorphic to $G_{SM}$ having the color $SU(3)$ in common. This justifies the interpretation given in Section 6 of the second factor $SU(3)$ above as the extended electroweak symmetry since ``by restriction" to the subalgebras $J_{(i)}$ corresponding to the 3 generations it leads to their corresponding electroweak symmetry ; this $SU(3)$ will be denoted by $SU(3)_{ew}$ so one can write $SU(3)_c\times SU(3)_{ew}/\mathbb Z_3$ for the subgroup of $F_4$ preserving the representation $\mathbb C\oplus \mathbb C^3$ of $\mathbb O$ as explained in Section 7.\\

Section 8 is our provisional conclusion.\\

The fact that the finite-dimensional quantum observable algebras are exactly the finite-dimensional Euclidean (i.e. formaly real) Jordan algebras is well understood and explained in the classical paper \cite{jor-vn-wig:1934}. The case of $J^8_3$ needs some care since this algebra is exceptional \cite{alb:1934} and therefore cannot enter within the usual Hilbert space framework ; a detailed analysis of this case has been given in \cite{gun-pir-rue:1978}.\\

The occurrence in particle physics of octonions and exceptional structures has a long history (see e.g. \cite{jor-vn-wig:1934}, \cite{gun-gur:1973}, \cite{gun-gur:1974}, \cite{gun-pir-rue:1978}, 
\cite{hor-bie:1979}, \cite{gur-tze:1996}, \cite{dix:1994}, \cite{ram:2003}, \cite{fur:2018}).\\

For the octonions, we refer to \cite{bae:2002}, for the exceptional Lie groups, we have used 
\cite{yok:2009} and for the Jordan algebras \cite{jac:1968}, \cite{sch:1995}, \cite{ber:2000}, \cite{mcc:2004}, \cite{ior:2009}, \cite{tow:2016}.\\

Concerning our notations, we use throughout the Einstein convention of summation on the repeated up-down indices. We denote by $\calm_n(A)$ the set of $n\times n$ matrices with entries in a $\ast$-algebra $A$ and by $\calh_n(A)$ the set of hermitian $n\times n$ matrices with entries in $A$ while the set of hermitian elements of $A$ will be simply denoted by $\calh(A)$.

\section{Unimodularity of $SU(3)_c$  and the quark-lepton symmetry}

The space of internal states of a quark is the 3-dimensional complex space $\mathbb C^3$ acted by the color group $SU(3)$. Since $SU(3)$  is a subgroup of $U(3)$ this means that the above state's space is a 3-dimensional complex Hilbert space $\mathbb C^3$ but the role of the unimodularity of $SU(3)$ that is of a corresponding normalized volume on $\mathbb C^3$ is not transparent. As an element of comparison, one has not the same problem with $\mathbb C^2$ acted by $SU(2)$ because since $SU(2)$ coincides with the group $U(1,\mathbb H)$ of 1-dimensional quaternionic unitaries it means that one has just to interpret $\mathbb C^2$ as a 1-dimensional quaternionic Hilbert space $\mathbb H$.\\

Now given the volume form $v$ on the complex Hilbert space $\mathbb C^3$ with $\vert\vert v\vert\vert=1$, one defines an internal antisymmetric antilinear cross-product $\times$ on $\mathbb C^3$ by setting
\begin{equation}
v(Z_1,Z_2,Z_3)= \langle Z_1 \times Z_2,Z_3\rangle
\end{equation}

for $Z_1,Z_2,Z_3\in \mathbb C^3$ where $\langle \centerdot, \centerdot\rangle$ is the sesquilinear scalar product of the Hilbert space $\mathbb C^3$. By chosing an orthonoral basis such that $v(Z_1,Z_2,Z_3)=\varepsilon_{k\ell m} Z^k_1 Z^\ell_2 Z^m_3$, one has in components
\begin{equation}
(Z_1\times Z_2)^k=\varepsilon_{k\ell m} \bar Z^\ell_1 \bar Z^m_2
\end{equation}
for the cross-product defined by (2.1). Thus one has now two products on $\mathbb C^3$, the internal cross-product $\times:\mathbb C^3\times \mathbb C^3\rightarrow \mathbb C^3$ and the scalar product $\langle \centerdot,\centerdot\rangle:\mathbb C^3\times \mathbb C^3\rightarrow \mathbb C$. Furthermore one has
\begin{equation}
\parallel Z_1\times Z_2\parallel^2=\parallel Z_1\parallel^2\parallel Z_2\parallel^2-\vert \langle Z_1,Z_2\rangle\vert^2
\end{equation}
for $Z_1,Z_2\in \mathbb C^3$. This suggests by adding a unit to define a product $\bullet$ on the Hilbert space direct sum $\mathbb C\oplus \mathbb C^3$ such that
\begin{equation}
\left\{
\begin{array}{c}
\parallel (z_1,Z_1)\bullet(z_2,Z_2)\parallel^2=\parallel (z_1,Z_1)\parallel^2 \parallel(z_2,Z_2)\parallel^2\\
\\
(0,Z_1)\bullet (0,Z_2)=(u\langle Z_1,Z_2\rangle^\sharp,Z_1\times Z_2)
\end{array}
\right.
\end{equation}
for $z_1,z_2 \in \mathbb C$ and $Z_1,Z_2\in \mathbb C^3$ where $u\in \mathbb C$ with $\vert u\vert=1$  and where the unit $\bbbone$ of $\mathbb C\oplus \mathbb C^3$ is $1\in \mathbb C\subset \mathbb C\oplus \mathbb C^3$. In (2.4), $\langle \centerdot,\centerdot\rangle^\sharp$ means either $\langle\centerdot,\centerdot\rangle$ or its complex conjugate $\overline{\langle\centerdot,\centerdot\rangle}$. Such a product is easy to construct, for instance by setting
\begin{equation}
(z,Z)\bullet (z',Z')=(zz'-\overline{\langle Z,Z'\rangle},zZ'+\bar z'Z+Z\times Z')
\end{equation}
for $z,z'\in \mathbb C$ and $Z,Z'\in \mathbb C^3$, which corresponds to the choice $u\langle Z_1,Z_2\rangle^\sharp=-\overline{\langle Z_1,Z_2\rangle}$. One has then
\begin{equation}
\overline{(z,Z)}\bullet (z,Z)=(z,Z)\bullet \overline{(z,Z)}=\parallel (z,Z)\parallel^2\bbbone
\end{equation}
by defining $\overline{(z,Z)}$ as $\overline{(z,Z)}=(\bar z,-Z)$. This implies that the underlying real algebra is a normed division algebra of real dimension 8 which is therefore isomorphic to the octonion algebra $\mathbb O$. The group $SU(3)$ is the group of complex linear transformation of $\mathbb C\oplus \mathbb C^3$ which preserves the product (2.5). We thus recover the classical fact that the subgroup of the automorphism group of $\mathbb O$ which preserves an imaginary octonionic unit (here the $i\in \mathbb C$) is isomorphic to $SU(3)$. Since the $SU(3)$-action on $\mathbb C\oplus \mathbb C^3$ is the fundamental action on $\mathbb C^3$, that is the action of the color group on the internal state's space $\mathbb C^3$ of a quark, and the trivial action on $\mathbb C$, it is natural to interpret $\mathbb C$ as the internal state's space of the corresponding lepton. In this way the quark-lepton symmetry corresponds to the splitting  of the octonion algebra $\mathbb O$ as $\mathbb C\oplus \mathbb C^3$ which itself is connected with the unimodularity of the color group $SU(3)$. This corresponds to the original approach of \cite{gun-gur:1973}.

\section{The 3 generations and the triality in $J^8_3$}

As recalled in Section 3.2 of \cite{mdv:2016b} and originally analyzed in \cite{jor-vn-wig:1934}, the algebra of observables of a finite quantum space should be a finite-dimensional Euclidean Jordan algebra which is clearly not the case of the octonion algebra $\mathbb O$.\\

However, as pointed out in Section 3.1 of \cite{mdv:2016b}, the existence of three generations of fundamental fermions reveals a sort of triality underlying the classification of these fundamental fermions. On the other hand, it is well known that the combination of $\mathbb O$ with the usual triality (of $Spin_8$) leads naturally to the exceptional Jordan algebra $J^8_3=\calh_3(\mathbb O)$ of hermitian octonionic $3\times 3$ matrices (see e.g. \cite{ada:1980}, \cite{jac:1968}, \cite{yok:2009}). This exceptional algebra is a perfect observable algebra for a finite quantum space which we call the exceptional quantum space, i.e. $J^8_3$ plays the role of the algebra of real functions on this (virtual) exceptional quantum space. This leads to the suggestion of \cite{mdv:2016b} that this finite quantum space corresponding to the exceptional Jordan algebra $J^8_3$ is relevant for the description of the internal state's space of the three generations of fundamental fermions.\\

We will follow this point of view here. But in order to understand the following of the paper, one observes that there are two apparently equivalent ways to describe the underlying triality of $J^8_3$ :
\begin{itemize}
\item
W1 - this triality corresponds to the three octonions of the matrice of an element of $J^8_3$,
\item
W2 - this triality corresponds to the three canonical subalgebras of hermitian octonionic $2\times 2$ matrices of $J^8_3$ corresponding themselves to the above three octonions of W1.
\end{itemize}

As in \cite{mdv:2016b}, this underlying triality of $J^8_3$ is interpreted as corresponding to the 3 generations. It turns out that, in spite of their obvious formal equivalence, the descriptions W1 and W2 lead naturally to 2 conceptually different interpretations.\\

If one adopts W1, as in \cite{mdv:2016b}, each octonion occurring in $J^8_3$ should correspond to one quark-lepton family but then remembering that one generation contains two families of quark-lepton having distinct sectors of charges, namely the neutrino and the electron families, and remembering that the $J^8_3$-modules are free we are led to consider a free module which is a factor of the sum of 2 copies $J^u$ and $J^d$ of $J^8_3$ with the particle assignment
\begin{equation}\label{AP1}
J^u=
\left(
\begin{array}{ccc}
\alpha_1 & \nu_\tau+t & \bar\nu_\mu-c\\
\bar \nu_\tau-t & \alpha_2 & \nu_e+u\\
\nu_\mu+c & \bar \nu_e-u & \alpha_3
\end{array}
\right), \>\>
J^d=
\left(
\begin{array}{ccc}
\beta_1 & \tau+b & \bar\mu-s\\
\bar \tau-b & \beta_2 & e+d\\
\mu+s & \bar e-d & \beta_3
\end{array}
\right)
\end{equation}
where the fundamental fermions are identified with their internal spaces that is $\mathbb C$ for the leptons and $\mathbb C^3$ for the quarks in the above representation $\mathbb C\oplus \mathbb C^3=\mathbb O$ of the octonion algebra, and where the $\alpha_k,\beta_k$ ($k\in \{1,2,3\}$) are new spin 1/2 fermions with $\mathbb R$ as internal space which do not carry any charge and should be described by Majorana spinors (see \cite{mdv:2016b}, Section 4).\\

If one adopts W2 which is our new approach here, the situation looks completely different because an algebra $J^8_2=\calh_2(\mathbb O)$ of hermitian octonionic $2\times 2$-matrices is an Euclidean Jordan algebra which is a spin factor
\begin{equation}\label{Sp9}
J^8_2=JSpin_9
\end{equation}
that  is a finite-dimensional quantum observable algebra corresponding to a finite quantum space can be associated to one complete generation  \cite{tod-dre:2018}.  More precisely, we shall prove in Section 4 that there is a canonical Euclidean extension of the observable algebra $J^8_2$ which corresponds to one full generation. Within this approach, the factor $J^8_3$ already potentially contains the 3 generations with their usual electroweak symmetries and without need to introduce new particles.\\

Let us remind the action of the subgroup $SU(3)\times SU(3)/\mathbb Z_3$ of the automorphism group $F_4$ of $J^8_3$ which preserves the representation of the octonions occurring in the elements of $J^8_3$ as elements of $\mathbb C\oplus \mathbb C^3$. For this,  one associates to the element
\begin{equation}
\left(
\begin{array}{ccc}\label{MJ}
\zeta_1 &x_3 & \bar x_2\\
\bar x_3 & \zeta_2 & x_1\\
x_2 & \bar x_1 & \zeta_3
\end{array}
\right)
\end{equation}
of $J^8_3=\calh_3(\mathbb O)$ the element of $J^2_3\oplus \calm_3(\mathbb C)= \calh_3(\mathbb C)\oplus\calm_3(\mathbb C)$
\begin{equation}\label{MC}
\left(
\begin{array}{ccc}
\zeta_1 &z_3 & \bar z_2\\
\bar z_3 & \zeta_2 & z_1\\
z_2 & \bar z_1 & \zeta_3
\end{array}
\right)
+ (Z_1,Z_2,Z_3)
\end{equation}
where
\begin{equation}\label{OC}
x_i=z_i+Z_i\in \mathbb C\oplus \mathbb C^3
\end{equation}
are the representation in $\mathbb C\oplus \mathbb C^3$ of the 3 elements of $x_i$ of $\mathbb O$. Then the action of $SU(3)\times SU(3)/\mathbb Z_3$ on $J^8_3$ is induced by the action of $SU(3)\times SU(3)$ on $\calh_3(\mathbb C)\oplus \calm_3(\mathbb C)$ given by
\begin{equation}\label{Act}
H\mapsto VHV^\ast,\>\> M\mapsto UMV^\ast
\end{equation}
for $(U,V)\in SU(3)\times SU(3)$ and $H\in \calh_3(\mathbb C)$, $M\in \calm_3(\mathbb C)$. Clearly the first $SU(3)$ factor of $SU(3)\times SU(3)/\mathbb Z_3(\subset F_4)$ is the color group.

\section{The factor $J^8_2=JSpin_9$ for one generation}

The spin factor $J^8_2=JSpin_9$ of hermitian octonionic $2\times 2$-matrices is a 10-dimensional Euclidean Jordan algebra which is strongly special \cite{jac:1968}. This means essentially that it is a special Jordan algebra which is such that its unital Jordan modules can be extracted from its universal unital associative envelop. The universal unital associative envelop of $J^8_2= JSpin_9$ is the Clifford algebra $C\ell(9,0)$.\\

The automorphism group of $J^8_2=JSpin_9$ is the group $Spin_9$ while the subgroup which preserves the splitting $\mathbb C\oplus \mathbb C^3$ of the octonion entering into the matrix of an element of $J^8_2$ is the subgroup 
\begin{equation}
G_{SM}=SU(3)\times SU(2) \times U(1)/\mathbb Z_6
\label{St}
\end{equation}
 of $Spin_9$ where the factor $SU(3)$ can be identified to the color group $SU(3)_c$. Indeed if one identifies $J^8_2$ with the Jordan subalgebra of $J^8_3$ consisting of the matrices of $J^8_3$ having vanishing first column and first row (which will be later associated to the first generation), then the above $G_{SM}$ coincides with the intersection in $F_4$ of the subgroup $SU(3)\times SU(3)/\mathbb Z_3$ with the subgroup $Spin_9$ of $F_4$ leaving invariant the idempotent $E_{(1)}\in  J^8_3$. It is worth noticing here that the subgroup of $Spin_9$ which preserves the splitting $\mathbb O=\mathbb C \oplus \mathbb C^3$ and acts by automorphism on $J^8_2= JSpin_9$ is in fact $U(3) \times SO(3)$.\\
 
 In order to understand this action of $U(3)\times SO(3)$ on $J^8_2$, one associates to the element
 \begin{equation}\label{j28}
 \left(
 \begin{array}{cc}
 \eta & x \\
 \bar x & \xi
 \end{array}
 \right)
 \end{equation}
 of $J^8_2$ the element
  \begin{equation}\label{jC}
 \left(
 \begin{array}{cc}
 \eta & z \\
 \bar z & \xi
 \end{array}
 \right) + Z
 \end{equation}
of $J^2_2\oplus \mathbb C^3=\calh_2(\mathbb C)\oplus \mathbb C^3$, where
\begin{equation}\label{deco}
x=z+Z\in \mathbb C\oplus \mathbb C^3
\end{equation}
is the representation of the octonion $x$ in $\mathbb C\oplus\mathbb C^3$. Then the action of\linebreak[4]
$U(3)\times SO(3)=U(3)\times (SU(2)/\mathbb Z_2)$ on $\calh_2(\mathbb C)\oplus \mathbb C^3$ is given by
\begin{equation}\label{act2}
H\mapsto VHV^\ast,\>\> Z\mapsto UZ
\end{equation}
 for $(U,V)\in U(3)\times SU(2)$ and $H\in \calh_2(\mathbb C)$, $Z\in \mathbb C^3$. Clearly the action of $U(3)$ corresponds to the action of $SU(3)_c\times U(1)$ on $\mathbb C^3$ while the action of $SO(3)$ corresponds to the adjoint action of $SU(2)$ on $\calh_2(\mathbb C)$.\\
 
The Clifford algebra $C\ell(9,0)$ has exactly 2 irreducible representations which are both of dimension $16=2^4$ (and which are inequivalent). The corresponding injective representation of $C\ell(9,0)$ is of dimension $32=2^5$. This is exactly {\sl after complexification} the dimension  of the internal space of the fundamental fermions for one generation. Later on in this section we will come back to the question of complexification.\\

This representation is for instance the one induced by the canonical identification and inclusion
\begin{equation}\label{even}
C\ell(9,0)=C\ell^0(9,1)\subset C\ell(9,1)
\end{equation}
where $C\ell^0(9,1)$ is the even part
of the Clifford algebra $C\ell(9,1)$ and where the isomorphisms 
\begin{equation}\label{C9.1}
C\ell(9,1)=\calm_{32}(\mathbb R)
\end{equation}
\begin{equation}\label{C09.1}
C\ell^0(9,1)=\calm_{16}(\mathbb R)\oplus \calm_{16}(\mathbb R)
\end{equation}
are classical ones.\\

It is worth noticing here that the structure group  \cite{jac:1976} $Str(J^8_2)$ of $J^8_2=JSpin_9$ is the direct product of 
\begin{equation}\label{St9}
Str_0(J^8_2)=Spin(9,1)
\end{equation}
with the multiplicative group of dilatations $\mathbb R^\times$, while one has 
\begin{equation}\label{Co9}
Co(J^8_2)=Spin(10,2)
\end{equation}
for the conformal (or M\"obius) group of $J^8_2=JSpin_9$.\\

Let us come back now to the question of complexification. Given a real unital special Jordan algebra $J$  with product denoted by $\circ$,  there are two natural universal unital associative envelops $S_1(J)$ and $S^c_1(J)$ for $J$. Namely $S_1(J)$ is the real associative unital algebra generated by $J$ with relations 
\begin{equation}\label{S-1}
\left\{
\begin{array}{l}
xy+yx=2x\circ y,\>\> \forall x, y\in J\\
\bbbone=\bbbone_J
\end{array}
\right.
\end{equation}
while $S^c_1(J)$ is the complex associative unital $\ast$-algebra generated by $J$ with relations
\begin{equation}\label{S-1-c}
\left\{
\begin{array}{l}
xy + yx = 2 x\circ y,\>\> \forall x,y\in J\\
\bbbone = \bbbone_J\\
x^\ast = x,\>\> \forall x\in J
\end{array}
\right.
\end{equation}
where $\bbbone$ is the unit of $S_1(J)$ or of $S^c_1(J)$ while $\bbbone_J$ is the unit of $J$.\\

The first definition $S_1(J)$ is the standard one in mathematics \cite{jac:1968} while the second one $S^c_1(J)$ is very natural for quantum theory (see e.g. \cite{mdv:2001},\cite{ac-mdv:2002a} and \cite{mdv-lan:2018}).\\

In the case where $J$ is a finite-dimensional special Euclidean Jordan algebra, $S^c_1(J)$ is 
 a finite-dimensional $C^\ast$-algebra and $\calh(S^c_1(J))$  is again an Euclidean Jordan algebra which is special and one has 
\begin{equation}\label{dS}
\dim \calh(S^c_1(J))=\dim (S_1(J))
\end{equation}
for these real algebras. Thus $\calh(S^c_1(J))$ is again a finite-dimensional quantum observable algebra which is a sort of completion of $J$.\\

In the case of $J^8_2=JSpin_9$, one has $S_1(J^8_2)=C\ell (9,0)$ while $S^c_1(J^8_2)$ will be denoted by $C\ell_9$. Thus $\calh(C\ell_9)$ is the {\sl completion} of $J^8_2$. The $C^\ast$-algebra $C\ell_9$ is given by 
\begin{equation}\label{Cl 9}
C\ell_9= \calm_{16}(\mathbb C)\oplus \calm_{16}(\mathbb C)
\end{equation}
which has a minimal injective representation as operator $C^\ast$-algebra in the Hilbert space
\begin{equation}\label{If1}
\mathbb C^{16}\oplus \mathbb C^{16}=\mathbb C^{32}
\end{equation}
that is the internal space of the fundamental fermions for one generation. The Euclidean Jordan algebra $\calh(C\ell_9)$ is given by
\begin{equation}\label{HCl9}
\calh(C\ell_9)=J^2_{16}\oplus J^2_{16}
\end{equation}
where $J^2_{16}=\calh_{16}(\mathbb C)$.\\

This gives the true basis for the realizations of 32-dimensional space of the fundamental fermions of one generation considered in \cite{tod-dre:2018}.

\section{W2-triality and the 3 generations}

We now consider the triality as corresponding to the 3 canonical Jordan subalgebras $J_{(1)}, J_{(2)}, J_{(3)}$ isomorphic to $J^8_2$ of $J^8_3$ where $J_{(1)}$ consists of the matrices of $J^8_3$
having vanishing elements in the first row and the first column, $J_{(2)}$ consists of the matrices having vanishing elements in the second row and the second column while $J_{(3)}$ consists of the matrices having vanishing elements in the third row and the third column. These 3 Jordan subalgebras $J_{(i)}$ correspond to the 3 diagonal primitive idempotent $E_{(i)}$ via 
\begin{equation}\label{JE}
J_{(i)}=(\bbbone - E_{(i)} ) J^8_3(\bbbone -E_{(i)})
\end{equation}
for $i\in \{1,2,3\}$.\\

These 3 subalgebras of $J^8_3$ isomorphic to $J^8_2$ will be associated to the 3 complete generations of fundamental fermions. $J_{(1)}$ being associated to the generation containing the leptons $e$ and $\nu_e$, $J_{(2)}$, being associated to the generation containing the leptons $\mu$ and $\nu_\mu$ and $J_{(3)}$ being associated to the generation containing the leptons $\tau$ and $\nu_\tau$.\\

It is interesting to localize the set of pure states $L_{(i)}$ of the $J_{(i)}(\simeq J^8_2)$ within the set of pure states $M$ of $J^8_3$.\\

The set $M$ is the Moufang plane that is the octonionic projective plane while the $L_{(i)}$ are projective lines in $M$ which intersect at the 3 primitive diagonal idempotents $E_{(j)}\in M$. Indeed $L_{(1)}$ is the projective line passing through $E_{(2)}$ and $E_{(3)}$, $L_{(2)}$ is the projective line passing through $E_{(3)}$ and $E_{(1)}$,  $L_{(3)}$ is the projective line passing through $E_{(1)}$ and $E_{(2)}$.\\

The subalgebras $J_{(i)}$ of $J^8_3$ are not independent and one has in $J^8_3$
\begin{equation}\label{InterJ}
J_{(i)} \cap J_{(j)}=\mathbb R E_{(k)}
\end{equation}
for any permutation $(i,j,k)$ of $(1,2,3)$. This corresponds to the intersections of the projective lines $L_{(i)}$ since one has canonically 
\begin{equation}\label {SJ}
M\subset J^8_3\>\> \text{and}\>\> L_{(i)}\subset J_{(i)}
\end{equation}
for $i\in \{1,2,3\}$.\\

It is worth noticing here that the $L_{(i)}$ do not include the pure states of fundamental fermions. The latter ones are primitive idempotents in the corresponding completed Euclidean Jordan algebras of the $J_{(i)}$.\\

Another aspect of the triality in $J^8_3$ is the existence in $F_4$ of the element $c$ defined by
\begin{equation}\label{c}
c\left(
\begin{array}{ccc}
\xi_1 & x_3 & \bar x_2\\
\bar x_3 & \xi_2 & x_1\\
x_2 & \bar x_1 & \xi_3
\end{array}
\right) =
\left(
\begin{array}{ccc}
\xi_2 & x_1 & \bar x_3\\
\bar x_1 & \xi_3 & x_2\\
x_3 & \bar x_2 & \xi_1
\end{array}
\right)
\end{equation}
which generates the cyclic group $\mathbb Z_3$ and satisfies
\begin{equation}\label{cJ}
\left\{
\begin{array}{l}
c(J_{(1)})=J_{(3)}\\
c(J_{(2)})=J_{(1)}\\
c(J_{(3)})=J_{(2)}
\end{array}
\right.
\end{equation}
that is
\begin{equation}\label{CJE}
c(J_{(i)})= J_{(i-1)},\>\> c(E_{(i)})=E_{(i-1)}
\end{equation}
where $i\in\{1,2,3\}\>\>  mod (3)$. One has
\begin{equation}\label{ac}
c(X)=aXa^{-1}
\end{equation}
with $a$ given by
\begin{equation}\label{a}
a=\left(
\begin{array}{ccc}
0 & 1 & 0\\
0 & 0 & 1\\
1 & 0 & 0
\end{array}
\right)
\end{equation}
and $X\in J^8_3$. Notice that $c$ is in the subgroup $SU(3)_c \times SU(3)/\mathbb Z_3$ of $F_4$ and corresponds to the action of the element $(1,a)$ of $SU(3)_c\times SU(3)$. It commutes with the action of $G_2$ and in particular with $SU(3)_c$. With the notation of the next section $a\in SU(3)_{ew}$.\\

In view of the triality, one shoud have 3 subgroups $Str_0(J_{(i)})$ of $Str_0(J^8_3)=E_{6(-26)}$ isomorphic to $Spin(9,1)$ and 3 subgroups $Co(J_{(i)})$ of $Co(J^8_3)=E_{7(-25)}$ isomorphic to $Spin(10,2)$ corresponding to the 3 canonical subalgebras $J_{(i)}$ of $J^8_3$ isomorphic to $J^8_2=JSpin_9$.

\section{The extended electroweak symmetry $SU(3)_{ew}$}

The subgroup of $F_4$ which preserves the splitting $\mathbb C\oplus \mathbb C^3$ of the octonions entering into the matrices of the elements of $J^8_3$ is 
\begin{equation}\label{Spl}
SU(3) \times SU(3)/\mathbb Z_3=SU(3)_c\times  SU(3)/\mathbb Z_3
\end{equation}
where the first factor is the color group $SU(3)_c$ as explained in \cite{mdv:2016b}. It is our aim in this section to give a similar interpretation for the second $SU(3)$ factor occurring in \ref{Spl}.\\

The automorphism groups of the Jordan algebras $J_{(i)}$ $(i\in \{1,2,3\})$ are 3 subgroups of $F_4=\Aut(J^8_3)$ isomorphic to $Spin_9$. The group $\Aut(J_{(i)})$ is also the subgroup of $F_4$ leaving invariant the idempotent $E_{(i)}$ for $i=1,2,3$.\\

The intersection in $F_4$ of $SU(3)_c\times SU(3)/\mathbb Z_3$ with $\Aut(J_{(i)})$ is a subgroup $G^{(i)}_{SM}$ isomorphic to the symmetry group of the Standard Model
\begin{equation}\label{StM}
G_{SM}=SU(3)_c\times SU(2) \times U(1)/\mathbb Z_6
\end{equation}
where the first $SU(3)$ factor is common for all $i \in \{1,2,3\}$ and is the color group $SU(3)_c$.\\

This means that the second $SU(3)$ factor projects for each of the 3 generations to its electroweak symmetry $U(2)$. It is why its natural interpretation is that of the {\sl extended electroweak symmetry} of the Standard Model with 3 generations and we denote it by $SU(3)_{ew}$ writing therefore $SU(3)_c\times SU(3)_{ew}/\mathbb Z_3$ for the symmetry group \ref{Spl}.

\section{The internal symmetry  $SU(3)_c\times SU(3)_{ew}/\mathbb Z_3$}

It is important to understand that in our approach here as well as in \cite{mdv:2016b}, the exceptional Jordan algebra $J^8_3$ plays the same role for the internal structure of fundamental particles as the algebra $\calc(M)$ of smooth functions on spacetime for the external structure.\\

In particular the automorphism group $F_4$ of $J^8_3$ plays the same role for the internal structure as the diffeomorphism group $\Diff(M)$ of spacetime for the external structure while the subgroup $SU(3)_c\times SU(3)_{ew}/\mathbb Z_3$ of $F_4$ plays for the internal structure the role of the Poincar\'e group (i.e the inhomogeneous Lorentz group) for the external structure.\\

This means that our internal space is the exceptional quantum space equipped with an additional structure (here the splitting $\mathbb C \oplus \mathbb C^3$ of $\mathbb O$), like the flat Lorentzian metric of the external spacetime.\\

\section{Discussion}

The advantage of the approach of this paper to the correspondence triality-generations with respect to the one of \cite{mdv:2016b} is that here one does not need the introduction of new fundamental fermions and that the formulation of the electroweak symmetry is straightforward. This new approach which is based on the key observation that the Euclidean Jordan algebra $J^8_2$ can be associated with a complete generation seems to be the most appealing one.\\

Nevertheless one should not forget that there is another possibility, namely the one suggested in Section 4 of \cite{mdv:2016b}, which is based on the same basic principles within this exceptional quantum geometry. Furthermore in this approach the interpretation of the group which preserves the splitting $\mathbb O=\mathbb C\oplus \mathbb C^3$ is straightforward (namely the quark-lepton symmetry).\\

In both cases one replaces the algebra of real functions on spacetime by the Jordan algebra of $J^8_3$-valued functions on spacetime which is interpreted as algebra of function on a corresponding ``almost classical quantum spacetime".\\

Since there is an additional structure involved on $J^8_3$ which is preserved by the subgroup $SU(3)_c\times SU(3)_{ew}/\mathbb Z_3$ of the automorphism group $F_4$ of $J^8_3$, the gauge theory giving the interactions will be based on $SU(3)_c\times SU(3)_{ew}/\mathbb Z_3$ as structure group with the gauge part $SU(3)_{ew}$ broken via the Higgs mechanism. It is expected that the corresponding Higgs fields are parts of the connection in the ``quantum directions".\\

There is an important step remaining which is to make the relation with the theory of Jordan modules in particular with the theory of $J_3^8$-modules. This is necessary for instance in order to use the differential calculus on Jordan algebras and Jordan modules developed in  \cite{mdv:2016b} and in \cite{car-dab-mdv:2018} to formulate the relevant dynamical theory and to get an interpretation of the Higgs sector. This is also of course important for the problems of classification.


\end{document}